\documentclass[english,aps,superscriptaddress,showkeys,showpacs,prepri,twocolumn]{revtex4}
\usepackage[T1]{fontenc}
\pdfoutput=1
\usepackage[letterpaper]{geometry}
\usepackage{wrapfig,rotating}
\usepackage{units}
\usepackage{bbding}
\usepackage{amsmath}
\usepackage{amssymb}
\usepackage{graphicx}
\usepackage{esint}
\usepackage[colorinlistoftodos,prependcaption,textsize=tiny]{todonotes}

\usepackage[utf8]{inputenc}
\usepackage{lmodern} 

\makeatletter

\@ifundefined{textcolor}{}
{%
 \definecolor{BLACK}{gray}{0}
 \definecolor{WHITE}{gray}{1}
 \definecolor{RED}{rgb}{1,0,0}
 \definecolor{GREEN}{rgb}{0,1,0}
 \definecolor{BLUE}{rgb}{0,0,1}
 \definecolor{CYAN}{cmyk}{1,0,0,0}
 \definecolor{MAGENTA}{cmyk}{0,1,0,0}
 \definecolor{YELLOW}{cmyk}{0,0,1,0}
}

\usepackage{doi}
\usepackage{hyperref}

\makeatother

\usepackage{babel}
\begin{document}

\title{Multidimensional tests of a finite-volume solver for MHD with a real-gas equation of state}

\author{J. R. King}
\affiliation{Tech-X Corporation, 5621 Arapahoe Ave. Boulder, CO 80303, USA}

\author{R. Masti}
\affiliation{Virginia Tech, Blacksburg, VA, USA}

\author{B. Srinivasan}
\affiliation{Virginia Tech, Blacksburg, VA, USA}

\author{K. Beckwith}
\affiliation{Sandia National Laboratories, Albuquerque, New Mexico, USA}


%
%
%

\date{draft \today}
\begin{abstract} 
This work considers two algorithms of a finite-volume solver for the MHD
equations with a real-gas equation of state (EOS). Both algorithms use a
multistate form of Harten-Lax-Van Leer approximate Riemann solver as
formulated for MHD discontinuities. This solver is modified to use the
generalized sound speed from the real-gas EOS. Two methods are tested: EOS
evaluation at cell centers and at flux interfaces where the former is more
computationally efficient. A battery of 1D and 2D tests are employed:
convergence of 1D and 2D linearized waves, shock tube Riemann problems, a 2D
nonlinear circularly polarized Alfv\'en wave, and a 2D
magneto-Rayleigh-Taylor instability test.  The cell-centered EOS evaluation
algorithm produces unresolvable thermodynamic inconsistencies in the
intermediate states leading to spurious solutions while the flux-interface
EOS evaluation algorithm robustly produces the correct solution. The
linearized wave tests show this inconsistency is associated with the
magnetosonic waves and the magneto-Rayleigh-Taylor instability test
demonstrates simulation results where the spurious solution leads to an
unphysical simulation.
\end{abstract}



\maketitle



  \newcommand{\vect}[1]{ \mathbf{#1}}
  \newcommand{\defn}{ \equiv}

  \newcommand{\lp}{\left(}
  \newcommand{\rp}{\right)}
  \newcommand{\lb}{\left[}
  \newcommand{\rb}{\right]}
  \newcommand{\la}{\left<}
  \newcommand{\ra}{\right>}

  \newcommand{\vf}{ \vect{f}}

  \newcommand{\vx}{\vect{x}}
  \newcommand{\vq}{\vect{q}}
  \newcommand{\vB}{\vect{B}}
  \newcommand{\vJ}{\vect{J}}
  \newcommand{\vA}{\vect{A}}
  \newcommand{\vE}{\vect{E}}
  \newcommand{\vV}{\vect{V}}
  \newcommand{\vF}{ \vect{F} }	
  \newcommand{\vU}{ \vect{U} }	
  \newcommand{\ddp}{\grad \cdot \Pi}
  \newcommand{\specheat}{\gamma_h}

  \newcommand{\grad}{\vect{\nabla}}
  \newcommand{\curl}[1]{\grad \times #1 }
  \newcommand{\dive}[1]{\grad \cdot #1 }
  \newcommand{\vdg}{\left(\vV \cdot \grad \right)}
  \newcommand{\bdg}{\left(\vB \cdot \grad \right)}
  \newcommand{\divV}{\grad \cdot \vV_1}
  \newcommand{\divVp}{\left( \grad \cdot \vV \right)}

  \newcommand{\dt}[1]{\frac{\partial #1}{\partial t}}
  \newcommand{\Dt}[1]{\frac{d #1}{dt}}
  \newcommand{\dpsi}[1]{\frac{\partial #1}{\partial \psi}}
  \newcommand{\dpsisq}[1]{\frac{\partial^2 #1}{\partial \psi^2}}
  
  \newcommand{\jac}{{\mathcal{J}}}
  \newcommand{\jaci}{{\mathcal{J}}^{-1}}
  \newcommand{\Pp}{ P^\prime }				
  \newcommand{\Vp}{V^\prime}
  \newcommand{\Vpp}{V^{\prime\prime}}
  \newcommand{\Vpo}{ \frac{V^\prime}{4 \pi^2}}
  \newcommand{\norm}{ P^\prime }		
  \newcommand{\RR}{ \psi }			
  \newcommand{\vR}{ \grad \RR }		
  \newcommand{\C}{ C }				
  \newcommand{\vC}{ \vect{\C} }		
  \newcommand{\vK}{ \vect{K} }			
  \newcommand{\vRsq}{ \mid \grad \RR \mid^2 }
  \newcommand{\vCsq}{ \C^2 }
  \newcommand{\vKsq}{ K^2 }
  \newcommand{\vBsq}{ B^2 }
  \newcommand{\vrr}{\frac{ \vR}{\vRsq} }
  \newcommand{\vbb}{\frac{ \vB}{B^2} }
  \newcommand{\vcc}{\frac{ \vC}{\vCsq} }
  \newcommand{\vjj}{\frac{ \vJ}{J^2} }
  \newcommand{\vkk}{\frac{ \vK}{\vKsq} }

  \newcommand{\R}{ \psi }
  \newcommand{\T}{ \Theta }
  \newcommand{\Z}{ \zeta }
  \newcommand{\A}{ \alpha }
  \newcommand{\U}{ u }
  \newcommand{\ve}{ \vect{e} }
  \newcommand{\vur}{ \vect{e}^\rho }
  \newcommand{\vut}{ \vect{e}^\Theta }
  \newcommand{\vuz}{ \vect{e}^\zeta }
  \newcommand{\vlr}{ \vect{e}_\rho }
  \newcommand{\vlt}{ \vect{e}_\Theta }
  \newcommand{\vlz}{ \vect{e}_\zeta }
  \newcommand{\gr}{ \grad \R }
  \newcommand{\gt}{ \grad \Theta }
  \newcommand{\gz}{ \grad \zeta }
  \newcommand{\ga}{ \grad \alpha }
  \newcommand{\gu}{ \grad \U }
  \newcommand{\dr}[1]{ \frac{\partial #1}{\partial \R} }
  \newcommand{\dT}[1]{\frac{\partial #1}{\partial \Theta}}
  \newcommand{\dz}[1]{\frac{\partial #1}{\partial \zeta}}
  \newcommand{\dU}[1]{\frac{\partial #1}{\partial \U}}
  \newcommand{\drs}[1]{ \frac{\partial^2 #1}{\partial \R^2} }
  \newcommand{\dTs}[1]{\frac{\partial^2 #1}{\partial \Theta^2}}
  \newcommand{\drt}[1]{\frac{\partial^2 #1}{\partial \R \partial \Theta}}
  \newcommand{\dzs}[1]{\frac{\partial^2 #1}{\partial \zeta^2}}
  \newcommand{\grr}{ g^{\R \R} }
  \newcommand{\grt}{ g^{\R \Theta} }
  \newcommand{\grz}{ g^{\R \zeta} }
  \newcommand{\gtz}{ g^{\Theta \zeta} }
  \newcommand{\gtt}{ g^{\Theta \Theta} }
  \newcommand{\gzz}{ g^{\zeta \zeta} } 
  \newcommand{\ri}{ \frac{1}{R^2} }
  \newcommand{\fr}{ \lp \R \rp}
  \newcommand{\frt}{ \lp \R, \T \rp}
  \newcommand{\frtz}{ \lp \R,\T,\Z \rp}

  \newcommand{\fluxav}[1]{\la #1 \ra}
  \newcommand{\thetaav}[1]{\la #1 \ra_\T}

\newcommand{\cramplist}{
        \setlength{\itemsep}{0in}
        \setlength{\partopsep}{0in}
        \setlength{\topsep}{0in}}
\newcommand{\cramp}{\setlength{\parskip}{.5\parskip}}
\newcommand{\zapspace}{\topsep=0pt\partopsep=0pt\itemsep=0pt\parskip=0pt}

\section{Introduction}
\label{sec:intro}

A common approximation for hydrodynamic simulations is to utilize
an equation of state (EOS) beyond of the simple `ideal gas' relation
and thereby model a so-called `real gas'.  With a real-gas EOS, the sound
waves are modified. This must be properly accounted for when using
numerical methods that are constructed via a flux-based description of the characteristic
waves of a PDE system, such as the finite volume method \cite{leveque} as studied
in this work. At the crux of this scheme is the reconstruction of an
approximate solution from a cell-centered location to the edge of a cell and
the construction of an approximate Riemann solution to a local nonlinear PDE
system to determine the flux at the discontinuous interface at the edge of a
cell. 

Numerical techniques for modeling a real gas composed of a neutral fluid are
well developed \cite{einfeldt1988}. For example, two techniques to accommodate
a real-gas EOS are direct modification of a Roe scheme \cite{roe81} and energy
relaxation. Within a Roe scheme, a linearized version of the conservative form
of the nonlinear system of equations is used to construct analytic eigenvalues
and eigenvectors that compose the flux at a given interface between two cells.
The appropriate form of the linearized flux Jacobian must be constructed with
an appropriate Roe average that properly accounts for the discontinuity in the
approximate solution at the cell interface and only admits physical shock
solutions but eliminates spurious solutions.  Unfortunately, the construction
of the traditional Roe average is predicated on the assumption of an ideal gas.  Buffard et
al., Ref.~\cite{Buffard2000}, describe the VFRoe scheme for use with the Euler
equations with a real-gas EOS.  This scheme uses the primitive variables in
the reconstruction with a simple arithmetic mean to determine the flux
Jacobian. A sonic entropy correction is used to eliminate nonphysical weak
solutions.  Alternatively, Coquel et al., Ref.~\cite{coquel98}, describe
energy-relaxation methods for a system of equations with a real-gas EOS.  In this
method the fluxes are computed from the ideal gas system with any chosen
Riemann solver and then an additional relaxation step is applied.  More
recently, Hu et al., Ref~\cite{Hu2009}, construct a generalized Roe average
that is appropriate for cases with a discontinuous EOS such as a material
interface. Hydrodynamic examples with a modified Harten-Lax Van Leer solver
\cite{HLL} that includes the contact discontinuity \cite{Toro1994} show that
this generalized Roe average is robust for multimaterial problems.

These techniques for real-gas systems are extended to magnetized and ionized
gases as described by the magnetohydrodynamic (MHD) equations by Dedner et al.,
in Ref.~\cite{Dedner2001}. The work of Dedner et al. only considers
one-dimensional test cases and both schemes (VFRoe adapted for MHD and a
relaxation method) work comparably well. Serna et al., Ref.~\cite{Serna2014},
consider the impact of a general EOS on the MHD system in terms of the
generation of anomalous wave structures produced by a non-convex system of
equations. One of the key results is that the anomalous wave structures can be
converted into classical wave structures when an appropriate amount of magnetic
field is applied. A number of one-dimensional cases with non-classical waves
without a magnetic field are tested with a modified version of the
characteristic-based nonconvex entropy-fix upwind scheme described in
Ref.~\cite{serna2009}.

In this work we focus on developing methods for real-gas applications that are
appropriately modeled by the MHD equations such as high-energy density
laboratory experiments. Given this application, the methods developed here are
intended for application to problems with more than one dimension.  We perform
a number of 1D tests to check convergence rates and to ensure results consistent
with Refs.~\cite{Dedner2001} and \cite{Serna2014} are achieved. However we also
extend our testing to include two dimensions as well.

This paper proceeds as follows: In Sec.~\ref{sec:analyticWaves} we review the
MHD system and the associated eigenvalue modification when a general EOS is
applied.  The approximate Riemann solvers tested in this work are of the
Harten-Lax Van Leer form in single-state \cite{Davis88}  and multiple-state \cite{miyoshi2005}
(including the contact and rotational Alfv\'en discontinuities) forms.  The solvers are
formulated in terms of the jump between the states at the interface and the
eigenvalues of the equation system as modified by a general EOS are used which
requires evaluation of the pressure and the generalized sound speed at the
interfaces between cells.  Sec.~\ref{sec:implementation} discusses the
specifics of two proposed algorithms: EOS evaluation at cell centers with
reconstruction to the cell interface or direct evaluation of the EOS tables at
the cell interface after the MUSCL reconstruction.  The latter method requires
more EOS evaluations and is thus more computationally expensive.  A series of
1D and 2D tests are performed with these algorithms in
Secs.~\ref{sec:1Dtests} and \ref{sec:2Dtests}, respectively. It is found that
although the cell-centered EOS evaluation works reasonably well for a number of
tests, only EOS evaluation at the interfaces is appropriate for high-fidelity
modeling.  Concluding remarks are made in Sec.~\ref{sec:summary}.

%

\section{Review of the MHD system with general EOS}
\label{sec:analyticWaves}

In conservative form, the ideal MHD equations with a real-gas EOS is given by
the continuity equation for the plasma mass density, $\rho$,
\begin{equation}
\frac{\partial\rho}{\partial t}+\nabla\cdot\left(\rho\mathbf{v}\right)=0\;,
\label{eq:continuity}
\end{equation}
the center-of-mass momentum, $\rho \mathbf{v}$, equation,
\begin{multline}
\frac{\partial\rho\mathbf{v}}{\partial t} = \\
-\nabla\cdot
\left(\rho\mathbf{v}\mathbf{v}+
      \left[P\left(\rho,E_{int}\right)+\frac{B^{2}}{2\mu_{0}}\right]\mathbf{I} 
-\frac{\mathbf{B}\mathbf{B}}{\mu_{0}}\right)\;,
\label{eq:momemtum}
\end{multline}
the induction equation for magnetic field, $\mathbf{B}$,
\begin{equation}
\frac{\partial\mathbf{B}}{\partial t}
-\nabla\times\left(\mathbf{v}\times\mathbf{B}\right)=0\;,
\label{eq:induction}
\end{equation}
and the energy, $E$, equation,
\begin{multline}
\frac{\partial E}{\partial t} = \\
-\nabla\cdot\left(\left[E+P\left(\rho,E_{int}\right)
      +\frac{B^{2}}{2\mu_{0}}\right]\mathbf{v}
+\left[\mathbf{v}\times\mathbf{B}\right]\times\mathbf{B}\right)\;.
\label{eq:energy}
\end{multline}
Here the total energy is defined as 
\begin{equation}
E=\frac{\rho v^{2}}{2}+\frac{B^{2}}{2\mu_{0}}+E_{int}\;,
\end{equation}
where $\mu_0$ is the permeability of free space, $E_{int}$ is the
internal energy, $P$ is the pressure and $\mathbf{I}$ is the 
identity tensor.

A constitutive relationship for $P(\rho,E_{int})$, otherwise known as an EOS,
is required to close the system of equations. The most common EOS is the
ideal-gas law,
\begin{equation}
P(\rho,E_{int})=\left(\Gamma-1\right)E_{int}\;,
\label{eq:idealEOS}
\end{equation}
where $\Gamma$, the ratio of specific heats, is a constant.  As
another analytic example, the Van Der Waals EOS is an extension that
qualitatively accounts for finite particle volume and intermolecular forces. 
The Van Der Waals EOS may be written as
\begin{equation}
P(\rho,E_{int})
=\frac{R}{C_{V}}\frac{E_{int}+\eta_{a}\rho^{2}}{1-\eta_{b}\rho}-\eta_{a}\rho^{2}
\label{eq:VDW}
\end{equation}
where $R$ is the gas constant, $C_{V}$ is the specific heat at constant volume,
$\eta_{a}$ a constant accounting for intermolecular forces and $\eta_{b}$ is a
constant accounting for molecule size.  More complicated EOS relations are
often not expressed analytically but rather are tabulated. For example, the
SESAME~\cite{SESAME} and Propaceos~\cite{PROPACEOS} EOS tables specify the
pressure and specific internal energy, $\epsilon = E_{int}/\rho$, in terms of
density and temperature. In practice, to use these tables with the equation set
of Eqns.~\eqref{eq:continuity}-\eqref{eq:energy} one must first invert
$\varepsilon\left(\rho,T\right)$ to solve for the temperature from a known
internal energy as an intermediate step, and then evaluate
$P\left(\rho,T\right)$.

Typically, a subset of the linear eigenvectors and eigenvalues of the EOS
system are required for approximate Riemann solvers. For
simplicity of exposition, we review the linear waves of the one-dimensional
Euler equations (essentially dropping the induction equation,
Eqn.~\eqref{eq:induction}, and $\mathbf{B}$ contributions). Consideration of
the 3D system only leads to additional, degenerate eigenvalues. The full
ideal-MHD system is considered in Ref.~\cite{Serna2014} with an analogous
modification of the sound speed.  Assuming Cartesian geometry and derivatives
only in the x-direction ($\partial/\partial x$ is denoted as a prime) leads to
the following system of equations:
\begin{equation}
\frac{\partial}{\partial t}\left(\begin{array}{c}
\rho\\
v_{x}\\
E
\end{array}\right)=\left(\begin{array}{c}
v_{x}\rho^{\prime}+v_{x}^{\prime}\rho\\
v_{x}v_{x}^{\prime}+P^{\prime}/\rho\\
v_{x}E^{\prime}+v_{x}P^{\prime}+\left(E+P\right)v_{x}^{\prime}
\end{array}\right)\;.
\end{equation}

In order to express this system in matrix form, we must eliminate
$P^{\prime}$. Noting that 
\begin{equation}
P^{\prime}\left(\rho,\varepsilon\right)
=P_{\rho}\rho^{\prime}+P_{\varepsilon}\varepsilon^{\prime}\;,
\end{equation}
and
\begin{equation}
E^{\prime}
=\rho v_{x}v_{x}^{\prime}+\left(\frac{v_{x}^{2}}{2}
+\varepsilon\right)\rho^{\prime}+\rho\varepsilon^{\prime}\;,
\end{equation}
we can solve this system to find
\begin{multline}
P^{\prime}\left(\rho,\varepsilon\right) = \\
\left[P_{\rho}-\frac{P_{\varepsilon}}{\rho}\left(\frac{v_{x}^{2}}{2}
+\varepsilon\right)\right]\rho^{\prime}-v_{x}P_{\varepsilon}v_{x}^{\prime}
+\frac{P_{\varepsilon}}{\rho}E^{\prime}\;.
\end{multline}
Here the subscripts of $P$ are used to denote partial derivatives (e.g.
$P_{\rho}=\partial P/\partial\rho$). This leads to the following characteristic
equations after reformulation as an eigenvalue problem:
\begin{widetext}
\begin{equation}
\det\left[\begin{array}{ccc}
v_{x}-\lambda & \rho & 0\\
\frac{1}{\rho}\left[P_{\rho}
-\frac{P_{\varepsilon}}{\rho}\left(\frac{v_{x}^{2}}{2}
+\varepsilon\right)\right] & 
v_{x}\left(1-\frac{P_{\varepsilon}}{\rho}\right)-\lambda &
\frac{P_{\varepsilon}}{\rho^{2}}\\
v_{x}\left[P_{\rho}-\frac{P_{\varepsilon}}{\rho}\left(\frac{v_{x}^{2}}{2}
+\varepsilon\right)\right] &
\left(E+P\right)-v_{x}^{2}P_{\varepsilon} &
v_{x}\left(1-\frac{P_{\varepsilon}}{\rho}\right)-\lambda
\end{array}\right]=0\;.
\end{equation}
\end{widetext}
Solving this system yields three eigenvalues, $\lambda=v_{x}+c_{s}$, $v_{x}$,
and $v_{x}-c_{s}$. The first and last of these are the sound waves and the
second is the entropy wave. The generalized expression for the sound speed is
then
\begin{equation}
c_{s}^{2}=\frac{P_{\varepsilon}P}{\rho^{2}}+P_{\rho}\;.\label{eq:GenSound}
\end{equation}
In the ideal gas limit, this leads to $c_{s}^{2}=\Gamma P/\rho$, as expected.

For the full MHD equation system,
Eqns.~\eqref{eq:continuity}-\eqref{eq:energy}, there are seven eigenvalues
after the divergence constraint on $\mathbf{B}$ is applied.  In particular, the
fast- and slow-magnetosonic waves can be written in terms of the sound speed and
these are modified with a general-EOS system where the sound speed is then
expressed as in Eqn.~\eqref{eq:GenSound}.  
\section{Algorithmic details}
\label{sec:implementation}

The specific implementation tested is the USim code \cite{USim} where the
full algorithmic details are described in this section.
A second or third order Runge-Kutta time discretization is used within
a finite volume scheme. In Sec.~\ref{sub:reconstruction} we describe
how fluxes at each interface are computed via a
monotonic upwind scheme for
conservation laws (MUSCL) reconstruction \cite{vanleer1974}. In particular a
total-variation diminishing (TVD), second-order accurate scheme that has been
shown to be robust for MHD in multi-dimensions \cite{stone2009} is used. 
Reconstruction of the conservative form of the
variables is tested within this work.  
After reconstruction to the interface, an approximate solution
to the Reimann problem is computed as described in Sec.~\ref{sub:reimann}.
Details on EOS computations are given in Sec.~\ref{sub:eosdetails}.
One complication that arises with a
multi-dimensional MHD equation system is the enforcement of the $\nabla \cdot
\mathbf{B}$ constraint \cite{toth2000}.  The application of this constraint is
trivial in one dimension. For multi-dimensional systems we use a hyperbolic
divergence cleaning scheme similar to that described in
Refs.~\cite{mignone2010,dedner2002}. 

\subsection{Second-order Accurate Reconstruction}
\label{sub:reconstruction}

\begin{figure}
  \includegraphics[width=6cm]{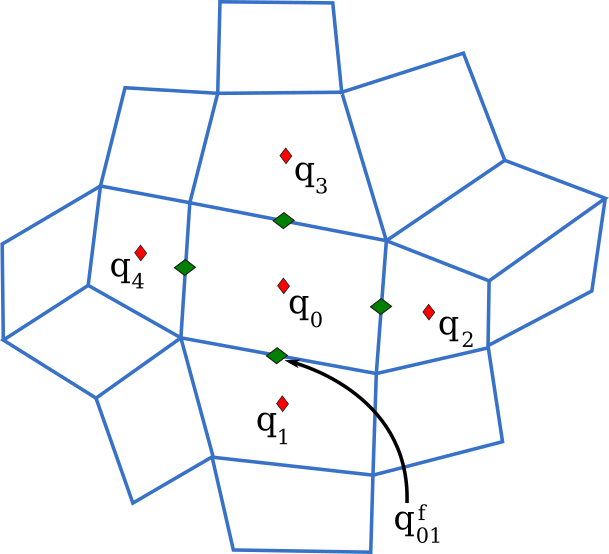}
  \caption{Stencil for reconstruction of fluid variables from cell- to
  face-centers.  USim computes a TVD-style second-order reconstruction of
  cell-centered data using a combination of the algorithms described in
  \cite{vanleer1974} and \cite{Mavriplis2003}.}
  \label{fig:gradientStencil}
\end{figure}

In order to achieve greater than first-order spatial accuracy within the MUSCL
scheme, fluid variables must be reconstructed from cell-centers to
cell-interfaces in a TVD fashion.  The USim variant of the MUSCL scheme
achieves second-order spatial accuracy using a variant of the original
\cite{vanleer1974} scheme, combined with the weighted least-squares gradient
scheme developed for compressible fluid dynamics on unstructured meshes
\cite{Mavriplis2003}.  Consider the quadrilateral mesh shown in
Figure~\ref{fig:gradientStencil} and some quantity $q_i$ located at cell
centers $i = 0,...,4$. If we denote $q$ at the face located between cells $0,1$
as $q^f_{01}$, then this quantity can be obtained in a TVD-fashion to
second-order accuracy as:
\begin{multline}
q^f_{01} = q_0 + \phi \left( r \right) \delta q; \\
\delta q = \left( 1 - w_0 \right) \left( q_1 - q_0 \right); \\
w_0 = \frac{ \left( \mathbf{x}_1 - \mathbf{x}_0 \right) 
        \cdot ( \mathbf{x}_1 - \mathbf{x}^f_{01} ) } 
      { \left| \mathbf{x}_1 - \mathbf{x}_0 \right|^2}
\end{multline}
Here, $\phi \left( r \right)$ is a TVD-style limiter, $r$ is a generalization
of the ratio of successive gradients suitable for unstructured meshes and
$\mathbf{x}_0, \mathbf{x}_1, \mathbf{x}^f_{01}$ are the coordinates of cell
$0$, cell $1$ and the face shared by these two cells, respectively. 
In this work the Van Leer limiter of Ref.~\cite{vanleer1974}, 
\begin{equation}
\phi \left( r \right) = \frac{r+|r|}{1+|r|} \;,
\end{equation}
is used.
In order to
provide a TVD-limited reconstruction, USim computes the ratio of successive
gradients as:
\begin{equation}
r = \frac{ \left( \mathbf{x}_1 - \mathbf{x}_0 \right) \cdot \nabla dq}{ 2 \delta q}
\end{equation}
where $\nabla dq$ is the weighted least-square gradient of $dq = q_i - q_0$
where the cells $i$ correspond to  the stencil of the weighted least-square
gradient operator described in the Appendix of \cite{Mavriplis2003}.

\subsection{Riemann Solvers}
\label{sub:reimann}

A version of the Harten-Lax-van Leer approximate Riemann solver
\cite{HLL} is used in the single-state (HLL, \cite{Davis88}) and multi-state 
(HLLD with `discontinuities', \cite{miyoshi2005}) forms.  These solvers apply
the Rankine-Hugoniot relations that specify the jump in states associated with
a subset of the waves of a Riemann problem. For an MHD system, these waves are
the fast and slow compressive shocks (two branches each), two branches of an
incompressible rotational discontinuity of the magnetic field and velocity
associated with the Alfv\'en wave, and an entropy wave involving only a jump in
the density and normal velocity.  HLL is a single-state solver that applies
jump conditions for the fast discontinuities whereas the HLLD
solver is a four-state solver that applies jump conditions for the fast,
Alfv\'en, and contact discontinuities.  

The HLL and HLLD approximate Riemann solvers are less diffusive than simpler
methods, such as Lax-Friedrichs, while being computationally cheaper than
linearized-flux methods that require computation of the full system of
eigenvectors (e.g. the methods of Roe \cite{roe81} or Serna
\cite{serna2009,Serna2014}).  For demonstration purposes we use a
Roe method with an arithmetic mean instead of the Roe
average as the Roe average is not well-defined for general EOS systems. As
demonstrated later, this leads to a ill-posed solution for some non-convex
problems. More advanced solvers such as the one proposed by Serna in
Ref.~\cite{Serna2014} overcome this limitation but are not tested in this work.

For the HLL and HLLD approximate Riemann solvers the maximum and minimum signal
speeds (eigenvalues) are defined in terms of the reconstructed left and right
interface states as in Ref.~\cite{Davis88}. Some formulations use a Roe average
in this definition~\cite{Einfelt91}, and using a form of the generalized Roe
average from Ref.~\cite{Hu2009} will be investigated in future work.  In
isolated cases where the HLLD approximate Riemann solver fails, the simple and
robust HLL formulation is used.

\subsection{EOS details}
\label{sub:eosdetails}

For all solvers the fast wave-speeds as modified by a general EOS are used
which requires evaluation of the pressure and the generalized sound speed at
the interfaces between cells. Two methods are explored for this evaluation: (1)
EOS evaluation at the cell centers with reconstruction to the cell interface or
(2) direct evaluation of the EOS tables after the MUSCL reconstruction. These
methods are more concisely referred to as the cell-centered (CC) and interface
(INTF) methods, respectively. The latter method requires more EOS evaluations
and is thus more computationally expensive. For example in 3D, where the ratio
of cell centers to cell edges is one to eight, we find the total execution
time takes 3.5x longer for the INTF method relative to the CC method. For this
timing test we use a spline fit to EOS tables with smoothing.

Evaluations of the generalized sound speed, Eqn.~\eqref{eq:GenSound}, requires
$P\left(\rho,\varepsilon\right)$ whereas the SESAME EOS tables provide
$P\left(\rho,T\right)$ and $\varepsilon\left(\rho,T\right)$. Thus in practice,
$\varepsilon\left(\rho,T\right)$ is inverted to find
$T\left(\rho,\varepsilon\right)$ which is then used to evaluate
$P\left(\rho,T\right)$. The inversion assumes that
$\varepsilon\left(\rho,T\right)$ is monotonic in $T$ and $\rho$. 

The partial derivatives, $P_{\varepsilon}$ and $P_{\rho}$, are not available
within the SESAME tables. Thus evaluation of Eqn.~\eqref{eq:GenSound}
requires some form of numerical differentiation if the tables are not
prescribed with an analytic form. The application of the chain rule requires
care as the tables typically provide $P\left(\rho,T\right)$ and
$\varepsilon\left(\rho,T\right)$ and not $P\left(\rho,\varepsilon\right)$.
Thus the partial derivative of pressure with respect to density at constant
specific energy is evaluated as 
\begin{multline}
P_{\rho}=\left.\frac{\partial P\left(\rho,\varepsilon\right)}{\partial\rho}\right|_{\varepsilon}=\frac{\partial P\left(\rho,T\right)}{\partial\rho}-\frac{\partial P\left(\rho,T\right)}{\partial\varepsilon}\frac{\partial\varepsilon}{\partial\rho}\\
=\frac{\partial P\left(\rho,T\right)}{\partial\rho}-\frac{\partial P\left(\rho,T\right)}{\partial T}\frac{\partial T}{\partial\varepsilon\left(\rho,T\right)}\frac{\partial\varepsilon\left(\rho,T\right)}{\partial\rho}\\
\simeq\frac{\partial P\left(\rho,T\right)}{\partial\rho}-\frac{\partial P\left(\rho,T\right)}{\partial T}\left[\frac{\partial\varepsilon\left(\rho,T\right)}{\partial T}\right]^{-1}\frac{\partial\varepsilon\left(\rho,T\right)}{\partial\rho}\;,
\end{multline}
where the last step assumes that $\varepsilon\left(\rho,T\right)$
is monotonic in $T$. Similarly, the partial derivative of pressure
with respect to specific energy at constant density is 
\begin{multline}
P_{\varepsilon}=
\left.\frac{\partial P\left(\rho,\varepsilon\right)}{\partial\varepsilon}\right|_{\rho}=
\frac{\partial P\left(\rho,T\right)}{\partial T}\frac{\partial T}{\partial\varepsilon\left(\rho,T\right)} \\
\simeq\frac{\partial P\left(\rho,T\right)}{\partial T}\left[\frac{\partial\varepsilon\left(\rho,T\right)}{\partial T}\right]^{-1}\;,
\end{multline}
where again we assume that $\varepsilon\left(\rho,T\right)$ is monotonic in $T$.
When analytic derivatives are not available, a
finite-difference method is used to evaluate these partial derivatives centered around
given values of $\rho$ and $T$ with a stencil size defined in terms of a small
$\delta$ (e.g.  $10^{-5})$ times $\rho$ and $T$.

Our experience is that a high-order fit to the table data that ensures
continuity of the derivative quantities is required to avoid spurious
oscillations. Thus for the SESAME results presented in this work, we use a
polynomial fit to the data within the region of interest. Another solution is
to use spline fits as may be provided through the EOSPAC library \cite{EOSPAC}.
We choose the simple polynomial fit as it is sufficiently accurate for our
purposes and it is computationally more efficient.
\section{1D test cases}
\label{sec:1Dtests}

We begin with a series of 1D tests that check the convergence rate of the
algorithm with a linearized wave test, Sec.~\ref{sec:lw1D}, and then compare to
prior results on Riemann problems, Sec.~\ref{sec:shockTube}.

\subsection{1D linear waves}
\label{sec:lw1D}

When a sufficiently small perturbation is placed upon large background fields
in a periodic box,
it permits analysis through linearization where terms proportional to the
square of the perturbation and higher order are dropped.  With constant
background fields the solution to this system of equations is simply the linear
waves.  In this test, a small perturbation proportional to the right
eigenvector of a given wave of the EOS system is used (see the appendix of
Ref.~\cite{Serna2014} for the right eigenvectors).  Perturbing a sinusoidal
wave using the right eigenvector associated with a particular characteristic
speed results in propagation of the wave at that speed.  A computation is run
for a single-wave period and the initial and final states are compared. A
difference in phase is attributed to dispersive error. Any reduction in wave
amplitude is due to diffusive error.  By varying the resolution, the calculated
L1 error in the eight MHD fields determine the convergence rate.

The vector of conserved variables, given by $\mathbf{Q}$, is perturbed using
the right eigenvectors associated with each of the slow-magnetosonic wave
$c_s$, the fast-magnetosonic wave $c_f$, and the Alfv\'en wave $c_a$.  The form
of the perturbation is
$\delta \mathbf{Q} = A\mathbf{R}\sin\left(2\pi \psi\right)$
where $A$ is the perturbation amplitude, $\mathbf{R}$ is the right eigenvector
associated with the wave of interest, and $\psi$ is $x/L_x$ for 1D, or
$x/L_x+y/L_y$ for 2D (see Sec.~\ref{sec:lw2D}). 

The SESAME table 5760 \cite{SESAME5760} is used for helium in this test.  The
initial conditions for these simulations include a density of $1.0$ $kg/m^3$, a
pressure of $6.13$ $atm$, and magnetic fields of $\mathbf{B_0}=619.4\; \langle 1.0,
\sqrt{2.0}, 0.5\rangle$ $T$.  This choice of magnetic fields maintains the same
plasma $\beta$ (the ratio of thermal pressure to magnetic pressure) as
\cite{Stone2008}. The perturbation amplitude is $A=10^{-6}$.  We study the
convergence of 1D simulations using a refinement factor of $2$ for grid
resolutions ranging from $32$ to $1024$ cells for a domain given by 
$0$ $m \leq x \leq1$ $m$.  
The time integration is a third-order
Runge-Kutta algorithm.

\begin{figure*}
  \includegraphics[width=15cm]{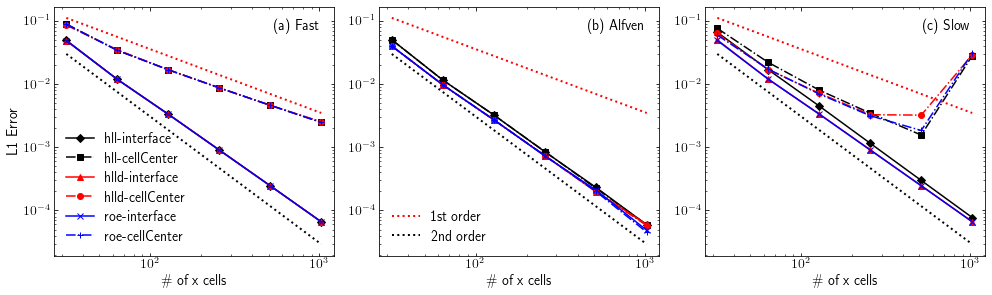}
  \caption{The 1D convergence of the interface- and cell-centered-EOS-evaluation
algorithms for the (a) fast magnetosonic, (b) Alfv\'enic, and (c) slow
magnetosonic perturbations, respectively. Separate curves are shown for the
HLL, HLLD and Roe approximate Riemann solvers.}
  \label{fig:lw1D}
\end{figure*}

Figure \ref{fig:lw1D} compares the L1 error for the linearized wave
computations for the three distinct waves, fast, Alfv\'en, and slow, at varying
resolution. The different algorithms for EOS evaluations, as described in
Sec.~\ref{sec:implementation}, are compared.  Finally, each algorithm is run
with the HLL, HLLD and Roe approximate Riemann solvers which creates six
curves per wave. In general, only small variations are evident between the
choice of approximate Riemann solver.

Both algorithms achieve the expected second-order convergence with the Alfv\'en
wave (middle plot).  This is reasonably expected as this wave is purely
magnetic and does not contain a pressure perturbation. However, deficiencies
with the cell-centered-EOS evaluations are evident with both the fast and slow
waves. For the fast wave, the cell-center-EOS-evaluation algorithm leads to
convergence at first order
while the interface-EOS-evaluation algorithm performs with
the expected second-order convergence. Similarly for slow wave while the interface-EOS-evaluation 
algorithm performs as expected, dispersion error ultimately leads to the lack
of convergence with the cell-centered-EOS-evaluation algorithm.  Visually, this
dispersion error is barely perceptible although it clearly is evident in the
plot of the L1 error.
\subsection{Shock tube}
\label{sec:shockTube}

\begin{table*}
  \begin{center}
  \begin{tabular}{l cccc cccc} \\[2mm] \hline
   & $\rho_L$ & $\mathbf{v}_L$ & $\mathbf{B}_L$ & $P_L$ 
   & $\rho_R$ & $\mathbf{v}_R$ & $\mathbf{B}_R$ & $P_R$ 
   \\[2mm] \hline
  Dedner01 & 250 & (0,0,0) & (61.752,54.277,0) & 35966778.0
           & 179 & (-1.7,5.6,0) & (61.752,49.413,0) & 26476136.8 \\[2mm]
  DG1      & 1.8181 & (0,0,0) & Tab.~\ref{tab:mag} & 3.000 
           & 0.275  & (0,0,0) & Tab.~\ref{tab:mag} & 0.575 \\[2mm]
  DG2      & 0.879  & (0,0,0) & Tab.~\ref{tab:mag} & 1.090 
           & 0.562  & (0,0,0) & Tab.~\ref{tab:mag} & 0.885 \\[2mm]
  DG3      & 0.879  & (0,0,0) & Tab.~\ref{tab:mag} & 1.090 
           & 0.275  & (0,0,0) & Tab.~\ref{tab:mag} & 0.575 \\[2mm]
  \hline
  \end{tabular}
  \caption{Initial conditions for the shock tube tests from
           Refs.~\cite{Dedner2001} and \cite{Serna2014}. Units are in
           $kg/m^3$, $m/s$, $mT$ and $Pa$, respectively, but are not 
           particularly meaningful for this problem. \label{tab:shock}}
  \end{center}
\end{table*}

\begin{table}
  \begin{center}
  \begin{tabular}{lcc} \\[2mm] \hline 
   & $\mathbf{B}_L$ & $\mathbf{B}_R$ \\[2mm] \hline
  MHD-DG1-O & (38.114,38.114,0) & same \\[2mm]
  MHD-DG1-T & (0,50.445,0) & same \\[2mm]
  MHD-DG2-O & (34.751,34.751,0) & same \\[2mm]
  MHD-DG2-T & (0,48.203,0) & same \\[2mm]
  MHD-DG3-O & (44.840,67.260,0) & same \\[2mm]
  MHD-DG3-T & (0,42.598,0) & same \\[2mm]
  MHD-DG3-R-O & (44.840,67.260,0) 
              & (44.840,-67.260,0) \\[2mm]
  MHD-DG3-L-O & (112.10,112.10,0) & same \\[2mm]
  \hline
  \end{tabular}
  \caption{Initial magnetic conditions for the shock tube tests from
           Ref.~\cite{Serna2014}. Units are in $mT$. \label{tab:mag}}
  \end{center}
\end{table}

\begin{table}
  \begin{center}
  \begin{tabular}{lcccc} \\[2mm] \hline 
    & $R$ & $C_V$ & $\eta_a$ & $\eta_b$ \\[2mm] \hline
  Dedner (Ref.~\cite{Dedner2001}) & 461.5 & 1401.88 & 1684.54 & 0.001692 \\[2mm]
  Serna (Ref.~\cite{Serna2014}) & 1 & 80 & 3 & $1/3$ \\[2mm]
  \hline
  \end{tabular}
  \caption{Van Der Waals model parameters for to characterize the EOS of 
           Eqn.~\eqref{eq:VDW} for the shock tube tests from
           Ref.~\cite{Serna2014}. \label{tab:vdw}}
  \end{center}
\end{table}

\begin{figure*}
  \includegraphics[width=15cm]{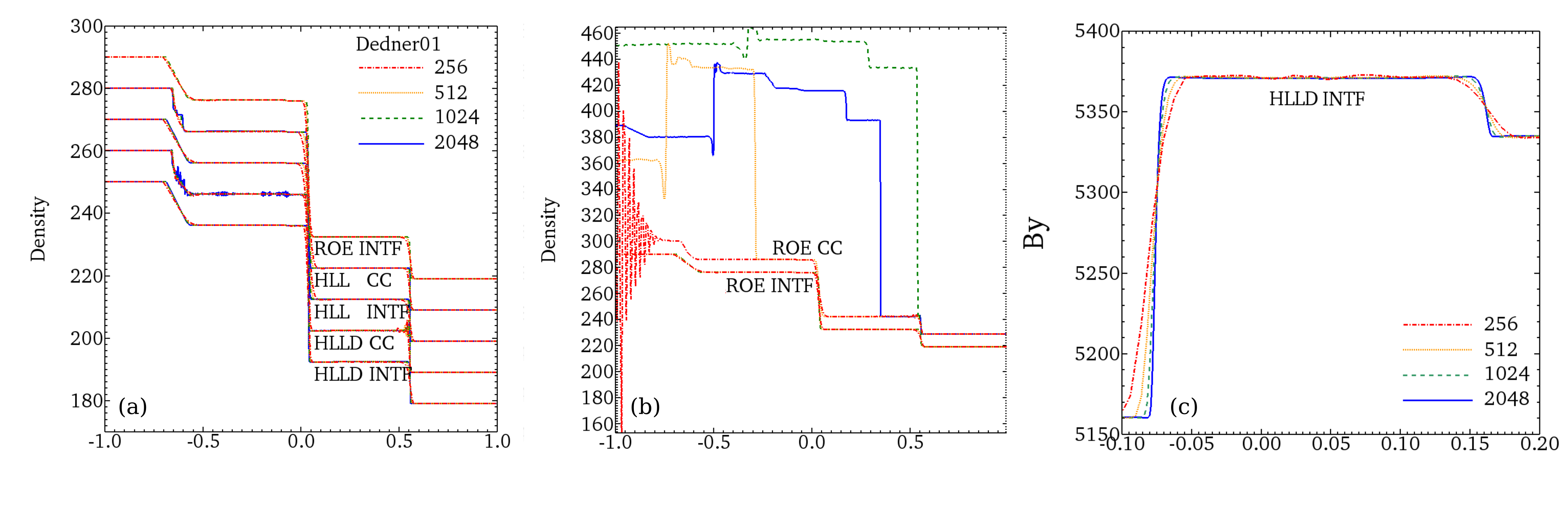}
  \caption{Six different algorithms compared to the results of the shocktube
test from Ref.~\cite{Dedner2001}, Fig.~6. In figures (a) and (b) the density
from a subset of five of the six algorithms and two of the six algorithms are shown,
respectively. The curves are grouped by resolution and are offset by multiples
of 10 $kg/m^3$ to allow for comparison.  The six different algorithms are the
combinations of the Roe with an arithmetic average, HLL and HLLD approximate
Riemann solvers and cell-centered- and interface-EOS-evaluation algorithms.  In figure (c)
the y-component of the magnetic field is plotted for the HLLD solver with the
interface-EOS-evaluation algorithm at different varied resolution. 
}
  \label{fig:dedner01}
\end{figure*}

In this section we compare computations with the known results of shock-tube
test cases from Refs.~\cite{Dedner2001} and \cite{Serna2014}. The initial
conditions for these cases are given in Tabs.~\ref{tab:shock} and
\ref{tab:mag}.  These cases use the Van Der Waals EOS with the parameters as
listed in Tab.~\ref{tab:vdw}. The time integration is a second-order
Runge-Kutta algorithm.

In Fig.~\ref{fig:dedner01} different methods are compared to the MHD
shocktube case shown in Fig.~6 of Ref.~\cite{Dedner2001}.  Six different
algorithms are compared: the combinations of the Roe with an arithmetic
average, HLL and HLLD approximate Riemann solvers and cell-centered- and
interface-EOS evaluation algorithms.  In the Fig.~\ref{fig:dedner01}(a), the density at
t=0.001 is plotted for all of these algorithms except Roe with
cell-centered-EOS evaluation.  The results are grouped by spatial resolution and
offset by
constant factors from the HLLD with interface EOS evaluation result to enable
quick comparison. The expected result from left to right is a compression fan
associated with the leftward propagating fast magnetosonic wave, a shock
associated with the leftward propagating slow magnetosonic wave (the jump in
the density is very small), a contact discontinuity, a shock associated with
the rightward propagating slow magnetosonic wave (the jump in the density is
again very small), and a shock associated with the rightward propagating fast
magnetosonic wave.  The interface-EOS-evaluation algorithm agrees well with the
expected results regardless of the choice of the approximate Riemann solver.
However, while the cell-centered-EOS-evaluation produces a somewhat reasonable
result at low resolution (except for an incorrect jump appearing in the
leftward compression fan that then appears as a compound wave), substantial
noise appears as the resolution is
increased. The appearance of this noise is associated with the magnetosonic waves
consistent with the dispersion error observed in the 1D linear wave tests in
Sec.~\ref{sec:lw1D}.  In Fig.~\ref{fig:dedner01}(b) the result for the Roe
solver with an arithmetic average with cell-centered EOS evaluation is also
shown. This algorithm produces
a clearly incorrect result, particularly at high resolution. In
Fig.~\ref{fig:dedner01}(c) the y-component of the magnetic field is plotted for
the HLLD solver with the interface EOS evaluation algorithm at different varied
resolution.  The convergence of this algorithm is favorable compared to the
algorithms tested in Ref.~\cite{Dedner2001} as seen by comparing to Fig.~6(b)
of that work.

\begin{figure*}
  \includegraphics[width=15cm]{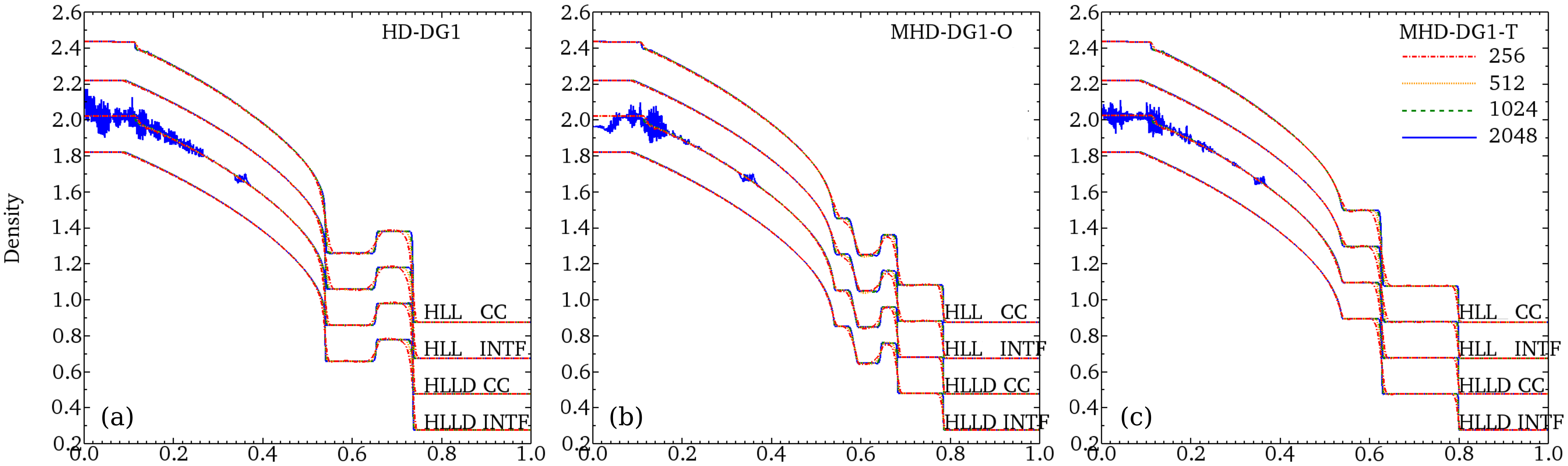}
  \caption{ Four different algorithms compared to the results of the DG1
  shocktube test at t=0.15 from Ref.~\cite{Serna2014}, Fig.~1.  The four
  different algorithms are the combinations of the HLL and HLLD approximate
  Riemann solvers and cell-centered- and interface-EOS-evaluation algorithms.  The density
  from the (a) hydrodynamic, (b) oblique-magnetic-field, and (c)
  transverse-magnetic-field versions of the tests are shown. The curves are
  grouped by resolution and are offset by multiples of 0.2 $kg/m^3$ to allow
  for comparison. }
  \label{fig:dg1}
\end{figure*}

\begin{figure*}
  \includegraphics[width=15cm]{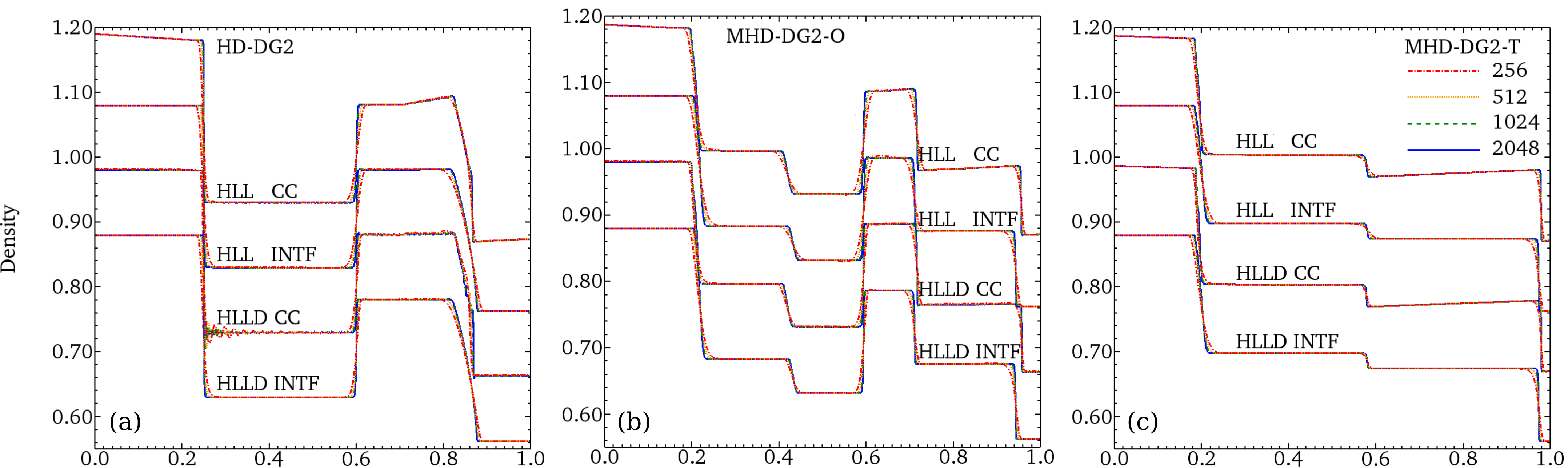}
  \caption{ Four different algorithms compared to the results of the DG2
  shocktube test at t=0.45 from Ref.~\cite{Serna2014}, Fig.~2.  The four
  different algorithms are the combinations of the HLL and HLLD approximate
  Riemann solvers and cell-centered- and interface-EOS-evaluation algorithms.  The density
  from the (a) hydrodynamic, (b) oblique-magnetic-field, and (c)
  transverse-magnetic-field versions of the tests are shown. The curves are
  grouped by resolution and are offset by multiples of 0.1 $kg/m^3$ to allow
  for comparison. }
  \label{fig:dg2}
\end{figure*}

\begin{figure*}
  \includegraphics[width=15cm]{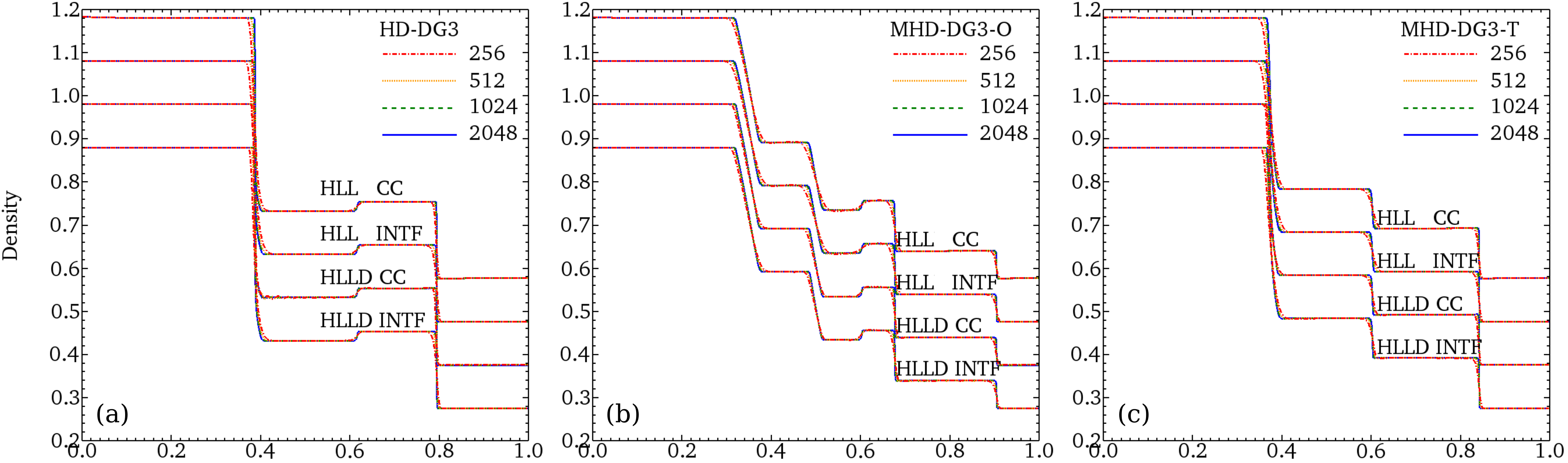}
  \caption{ Four different algorithms compared to the results of the DG3
  shocktube test at t=0.2 from Ref.~\cite{Serna2014}, Fig.~3.  The four
  different algorithms are the combinations of the HLL and HLLD approximate
  Riemann solvers and cell-centered- and interface-EOS-evaluation algorithms.  The density
  from the (a) hydrodynamic, (b) oblique-magnetic-field, and (c)
  transverse-magnetic-field versions of the tests are shown. The curves are
  grouped by resolution and are offset by multiples of 0.1 $kg/m^3$ to allow
  for comparison. }
  \label{fig:dg3}
\end{figure*}

Next we compare results from the shocktube tests of Ref.~\cite{Serna2014} in
Figs.~\ref{fig:dg1}-\ref{fig:dg3}. These should be compared to Figs.~3-5 of
Ref.~\cite{Serna2014}.  Each of these tests start with a pure hydrodynamic case
(HD) without magnetic field. Evaluation of the thermodynamic fundamental
derivative shows that all of these cases produce non-convex dynamics with
anomalous waves for the hydrodynamic cases \cite{Serna2014}.
As such, the ability of the scheme to produce solutions comparable with those
of Ref.~\cite{Serna2014} is a useful indicator of its ability to
address non-convex EOS in a robust fashion; such a capability is critical for
algorithms that address real materials, due to the presence of non-convex
behavior within tabular EOS. Furthermore, the presence of oblique (MHD-O) and
transverse (MHD-T) magnetic field within the test modifies the overall
convexity of the system. In the cases presented here, the presence of these
magnetic fields act to increase the convexity of the overall system; accurately
capturing this interplay is important to understanding the interaction of
magnetic fields with real materials.

In Figs.~\ref{fig:dg1}, \ref{fig:dg2} and \ref{fig:dg3} the results of the
DG1-DG3 cases are shown. For all of these cases, the Roe solver
with an arithmetic average fails to produce a numerical solution and we 
do not include it in this discussion. From left to right for the HD case, there is a
expansion shock or rarefaction fan, an entropy wave and compression shock or
fan. For the DG1 and DG2 cases, Figs.~\ref{fig:dg1} and \ref{fig:dg2}
respectively, the cell-centered-EOS-evaluation algorithm produces an errant
feature at the top of the leftmost expansion and with the relatively low
dissipation HLLD solver spurious oscillations are observed at high resolution.
All algorithms perform well on the DG3 case in Fig.~\ref{fig:dg3} (this is also
true for the rotated-oblique-magnetic-field and large-oblique-magnetic-field
versions of DG3 from Ref.~\cite{Serna2014} which are not shown).  With the
addition of oblique (MHD-O) and transverse (MHD-T) magnetic field
additional waves are produced as associated with the MHD system. The result of
comparison of the cell-centered- and interface-EOS-evaluation algorithms is
comparable to that found with the HD results.


\section{2D test cases}
\label{sec:2Dtests}

Next we apply our algorithm to 2D systems to check the performance with
multiple dimensions. The tests progress in order of increasing complexity:
first a linear wave test of convergence, Sec.~\ref{sec:lw2D}, then the
nonlinear circularly-polarized Alfven wave test, Sec.~\ref{sec:cp}, which
checks both convergence for a problem with a large amplitude perturbation but
also tests the accuracy of the algorithm.  Finally, we conclude with a test
using the magnetized Rayleigh-Taylor instability, Sec.~\ref{sec:rt}, which
fully demonstrates the limitations of the cell-centered reconstruction
algorithm.

\subsection{2D linearized waves}
\label{sec:lw2D}

The two-dimensional linearized-waves tests are an extension of the 1D tests
from Sec.~\ref{sec:lw1D}.  In 2D, the vector components of $\mathbf{Q}$ undergo
a rotational transformation to create a propagation oblique to the grid.  The
2D simulations use a structured, Cartesian domain ($0$ $m\leq x \leq2$ $m$ and $0$
$m\leq y \leq1$ $m$) with twice the grid resolution in the x-direction compared
to the y-direction. Since the grid is rectangular, and the cells are square,
the transformed perturbation along the rectangular diagonal ensures that the x
and y fluxes differ in magnitude for each cell; thereby ensuring the simulation
is truly 2D.  The convergence tests start with a grid resolution in the
y-direction of $8$ cells and is ultimately enhanced to $256$. 

\begin{figure*}
  \includegraphics[width=15cm]{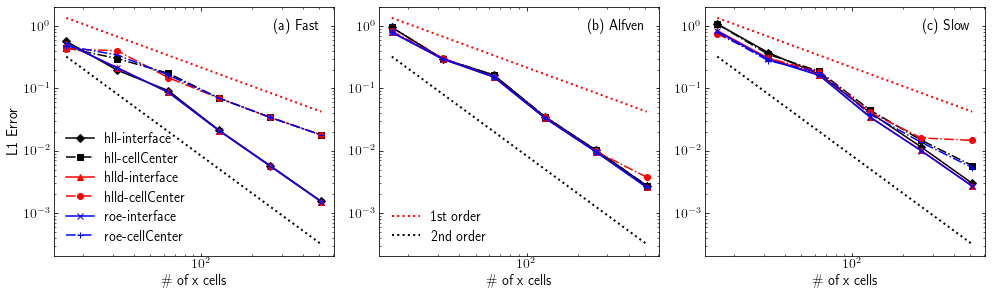}
  \caption{The 2D convergence of the interface- and cell-centered-EOS-evaluation
algorithms for the (a) fast magnetosonic, (b) Alfv\'enic, and (c) slow
magnetosonic perturbations, respectively. Sepearate curves are shown for the
HLL, HLLD and Roe approximate Riemann solvers.}
  \label{fig:lw2D}
\end{figure*}

Figure~\ref{fig:lw2D} shows the convergence of the interface- and cell-centered-
EOS-evaluation algorithms for the (a) fast-magnetosonic, (b) Alfv\'enic, and (c)
slow-magnetosonic perturbations, respectively. Separate curves are shown for
the HLL, HLLD and Roe approximate Riemann solvers.  The convergence in 2D is
comparable to the 1D result from Fig.~\ref{fig:lw1D}.  For the interface-EOS-evaluation 
algorithm, the expected second-order convergence for all waves is
produced. However, the results for the cell-centered-EOS-evaluation algorithm
are mixed: Although the Alfv\'en wave converges with second-order accuracy,
the fast-magnetosonic wave converges only at first order. The slow-magnetosonic 
wave initially converges with second-order accuracy but it ultimately
produces large errors at high resolution resulting from dispersion.
Again this result shows the importance of consistent EOS evaluation for the
compressive magnetosonic waves.
\subsection{Nonlinear 2D circularly polarized Alfv\'en wave}
\label{sec:cp}

Next we examine a nonlinear circularly-polarized Alfv\'en wave test which checks
both convergence for a problem with a large amplitude perturbation but also
tests the accuracy of the algorithm. This test is based off a version from
Ref.~\cite{toth2000} but is modified to the grid and wave orientation from
Ref.~\cite{Gardiner2005}. In this form, the initial state is $\rho=1$,
$p=\beta b^2/2$,
\begin{align}
\begin{split}
  \rho u_x &= -0.1 sin(\alpha) sin(2\pi x_\parallel) \; kg \, m^{-2} \, s^{-1}\\ 
  \rho u_y &=  0.1 cos(\alpha) sin(2\pi x_\parallel) \; kg \, m^{-2} \, s^{-1}\\
  \rho u_z &=  0.1 cos(2\pi x_\parallel) \; kg \, m^{-2} \, s^{-1}\\
       B_x &=  cos(\alpha) - 0.1 sin(\alpha) sin(2\pi x_\parallel) \; \sqrt{\mu_0} \, T\\
       B_y &=  sin(\alpha) + 0.1 cos(\alpha) sin(2\pi x_\parallel) \; \sqrt{\mu_0} \, T\\
       B_z &=  0.1 cos(2\pi x_\parallel) \; \sqrt{\mu_0} \, T.
\end{split}
\end{align}
where the parallel direction to the wavevector is given by $x_\parallel = x
cos(\alpha) + y sin(\alpha)$.  This leads to
a constant magnetic-field energy of $0.505$ $J/m^3$.  The domain is zero to
$1/cos(\alpha)$ in the x direction and zero to $1/sin(\alpha)$ in the y
direction with $\alpha=\pi/2-atan(0.5)\simeq63.4^{\circ}$.
The number of x cells is twice the y cells such that each cell is then square
and propagation is oblique to the grid. The time integration is a third-order
Runge-Kutta algorithm.  The Van Der Waals (Eqn.~\eqref{eq:VDW}) model
parameters for the non-ideal EOS computations are $R=2.077$, $C_V=3.116$,
$\eta_a=9.213\times10^{-4}$ and $\eta_b=9.513\times10^{-2}$. These parameters
are chosen such that there is an approximately 10\% variation to the EOS
relative to the ideal-gas EOS.

\begin{figure*}
  \includegraphics[width=12.5cm]{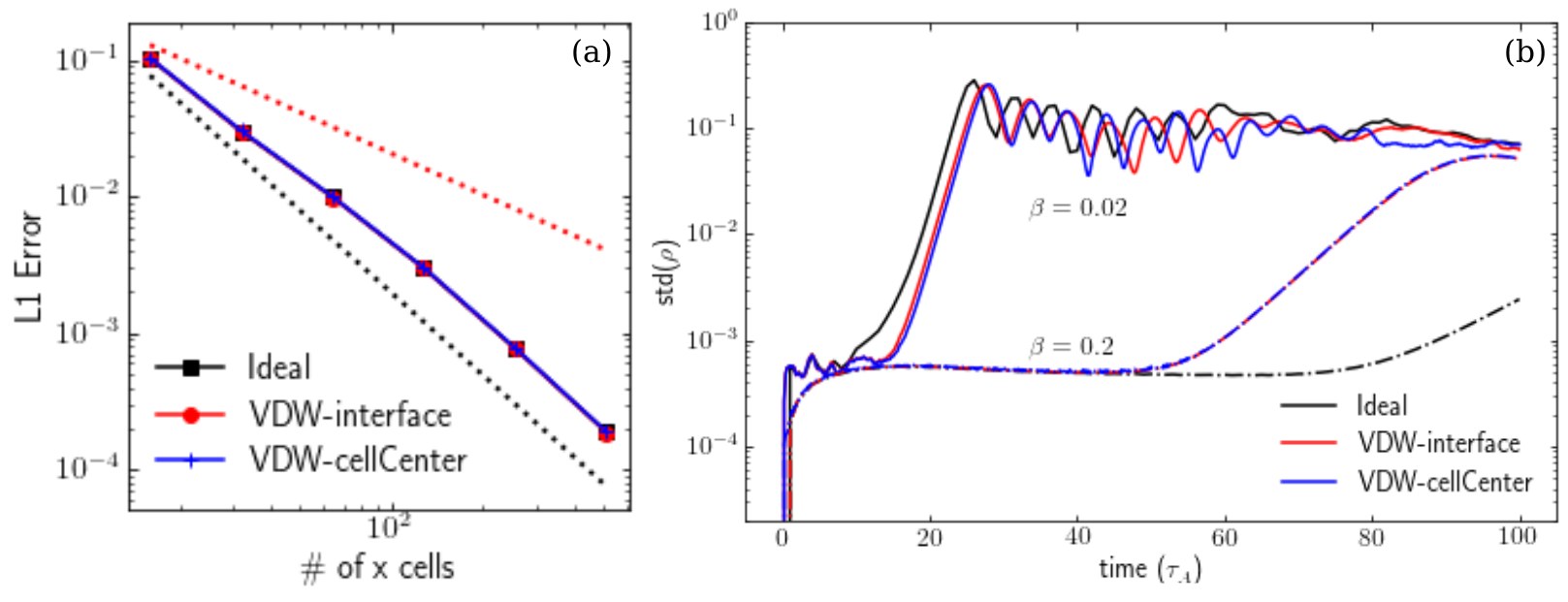}
  \caption{Results from the circularly polarized Alfv\'en wave test.  Subfigure
  (a) shows the L1 error in the momentum after a single period of oscillation
  of the wave for the ideal model and the Van Der Waal model with the
  cell-centered- and interface-EOS-evaluation algorithms. Subfigure (b) shows
  the standard deviation of the density with respect to time for the three
  algorithms/models at two different values of $\beta$.}
  \label{fig:cp}
\end{figure*}

Figure \ref{fig:cp}(a) plots the L1 error in the momentum after a single period
of oscillation of the wave for the ideal model and the Van Der Waal model with
the cell-centered- and interface-EOS-evaluation algorithms. Convergence of the
algorithm is second order for all cases consistent with the discussions of 
Secs.~\ref{sec:lw1D} and \ref{sec:lw2D}.

As described in Ref.~\cite{DelZanna2001} it is known that the
circularly-polarized Alfv\'en wave is subject to a parametric instability that
causes it to decay into a magnetosonic wave.  The rate of decay is inversely
proportional to the plasma $\beta=2\mu_0p/B^2$. Similar to Ref.~\cite{DelZanna2001}
we examine the time dynamics of this parametric decay in simulations at two
values of $\beta$ and varied numerical algorithms and/or models: the HLLD
approximate Riemann solver with an ideal-gas EOS and a Van Der Waals EOS with
either interface- or cell-centered-EOS evaluations. Unlike
Ref.~\cite{DelZanna2001}, we do not seed our simulation with random noise, thus
we expect the parametric instability to rise purely from numerical error.  

Fig.~\ref{fig:cp}(b) shows the standard deviation of the density with respect
to time for the three algorithms/models at two different values of $\beta$ for
a fixed grid resolution of 128$\times$64.  A large standard deviation of the
density indicates that compressional magnetosonic dynamics have become large
with the simulation entering a turbulent state. As the growth rate is inversely
proportional to $\beta$, at $\beta=0.02$ the instability grows faster than at
$\beta=0.2$. For both cases the sound speed is approximately 5\% greater for
the Van Der Waals EOS ($c_s=0.136$ m/s) relative to the ideal gas EOS
($c_s=0.13$ m/s). For both values of $\beta$, the cell-centered EOS evaluation
algorithm performs nearly identical to the interface EOS evaluation algorithm
until well after a saturated state is achieved. As the instability is seeded by
numerical error, this reinforces the results of Sec.~\ref{sec:lw1D} and
\ref{sec:lw2D} in that the error associated with the Alfv\'en wave is
comparable between these algorithms.
\subsection{Magneto-Rayleigh-Taylor instability}
\label{sec:rt}

Finally, we examine the performance of the EOS algorithms with the classical
single-mode Magnetic Rayleigh-Taylor (MRT) instability simulation from
Ref.~\cite{Stone2008}. Our computations differ from the prior work as
more realistic densities, pressures, and temperatures for the ideal gas helium
are used with the SESAME-5760 EOS similar to  Secs.~\ref{sec:lw1D} and
\ref{sec:lw2D}.  In this test, a heavy plasma is supported by a light plasma at
constant pressure in the presence of a gravitational acceleration and a
transverse magnetic field. This configuration is an inherently unstable
equilibrium, and the vertical velocity in the direction of the acceleration is
sinusoidally perturbed to seed the instability. 

Initial conditions for this simulation are $\rho_h=5.0$ kg/m$^3$,
$\rho_l=0.5\rho_h$, $P_0=2$ atm, $g=24.0$ km/s$^2$, and
$\vec{B}=\langle25.0,\,0.0,\, 0.0\rangle\;mT$ which correspond to heavy-fluid
density, light-fluid density, atmospheric pressure, gravitational acceleration,
and the magnetic field, respectively. The constant $g$ is chosen such that the
buoyancy pressure ($=\rho g y$) is $P_b\approx 0.06 P_0$ ensuring the MRT
interface, the boundary between the heavy and light fluid, remains relatively
fixed, and coincidentally the gage pressure $P=P_0-P_b=P_0-\rho g y$ does not
go negative while still allowing for a reasonable growth rate given the spatial
scales.  The magnetic field is chosen so that the simulation is spatially
converged late in time, and not overly stabilized. Without a magnetic field the
numerical error from the development of mesh-scale features outweighs any
different dynamics introduced from using a real-gas EOS which makes comparison
difficult. 

The domain is $L_x=1$ $m$, and the range is $L_y=2$ $m$, with a grid resolution
of 200$\times$400 where the origin is chosen such that
$x\epsilon[-L_x/2,L_x/2]$, and $y\epsilon[-L_y/2,L_y/2]$. Boundary conditions
for the y-direction are non penetrating, and for the x-direction are periodic.
The time integration is a second-order Runge-Kutta algorithm.

The widely known classical linear growth rate of the MRT instability is given
by
\begin{equation}
  \gamma = \sqrt{k g A - \frac{(\vec{k}\cdot\vec{B})^2}{\mu_0(\rho_h+\rho_l)}}
\end{equation}
where $\vec{k}$ is the perturbation wavevector which has magnitude $k=2\pi$
$m^{-1}$ and $A$ is known as the Atwood number which is
$A=(\rho_h-\rho_l)/(\rho_h+\rho_l)=1/3$. For the test conditions $\gamma
\approx 218$ $s^{-1}$ which is used to calculate an simulation time of
$6.0/\gamma\approx 27.5$ $ms$. This time allows for sufficient mode growth well
into the nonlinear regime. The derivation of the linear growth rate does not
assume an ideal-gas EOS, and this allows direct comparison of MRT growth
between the different EOS algorithms. The functional form of the
perturbation on the vertical velocity is a single sinusoid with $\lambda$
equivalent to the domain size $L_x$, and is given by
\begin{equation}
\delta v = -\delta\left(1+\cos(k x)\right)\exp\left[\frac{-20 y^2}{L_y^2}\right]\;,
\end{equation}
where $\delta$ is the perturbation amplitude which is $6.6$ $m/s$. 

\begin{figure}
  \includegraphics[width=1.05\linewidth]{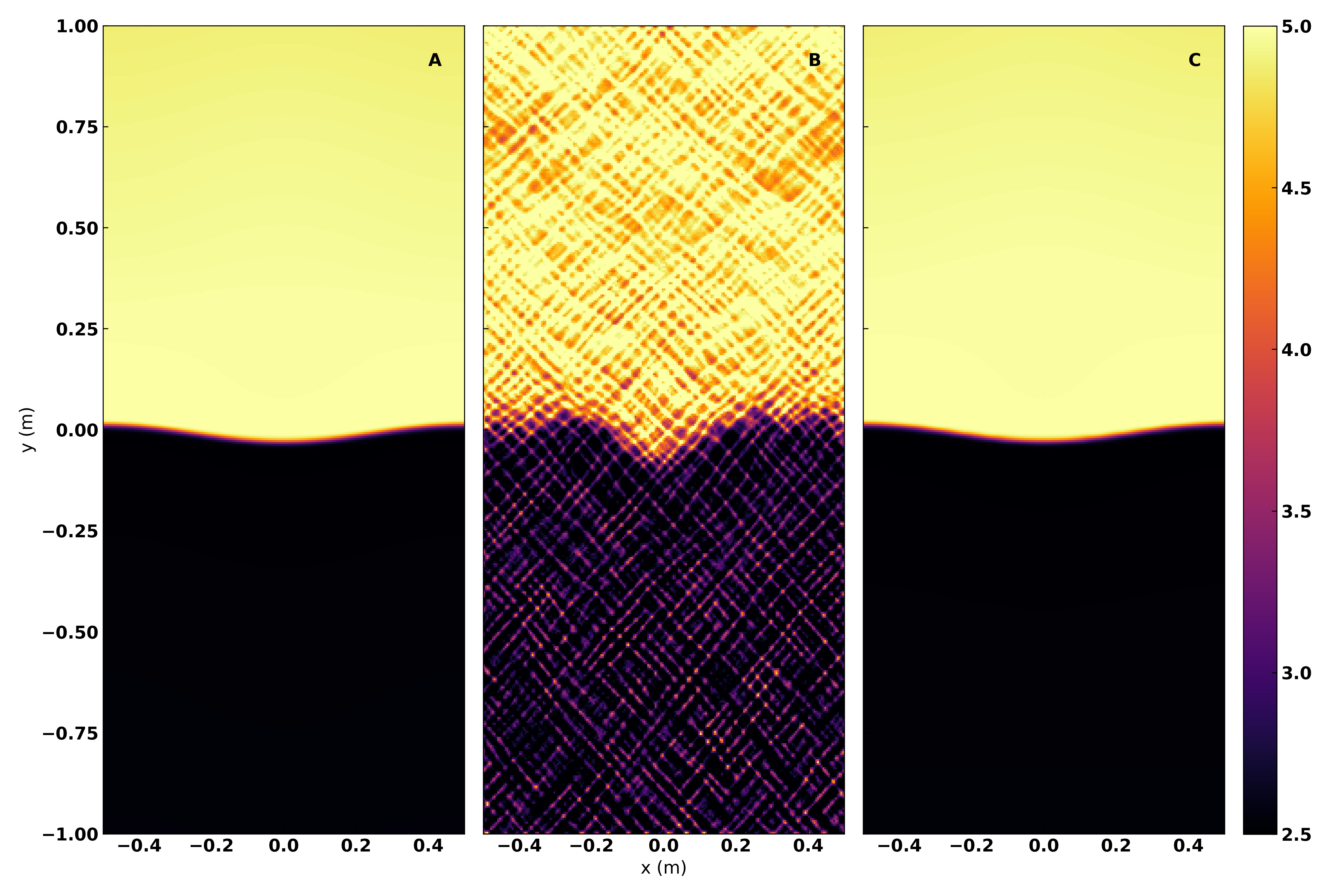}
  \caption{The early-time density state in kg/m$^3$ ($t= 3.3\;ms$) of the 2D
  MRT instability growth of density in kg/m$^3$ for the (A) ideal-gas EOS, (B)
  the SESAME-5760 EOS using the cell-centered EOS evaluation algorithm, and (C)
  the SESAME 5760-EOS using the interface EOS evaluation algorithm.}
  \label{fig:rtEarly}
\end{figure}

\begin{figure}
  \includegraphics[width=1.05\linewidth]{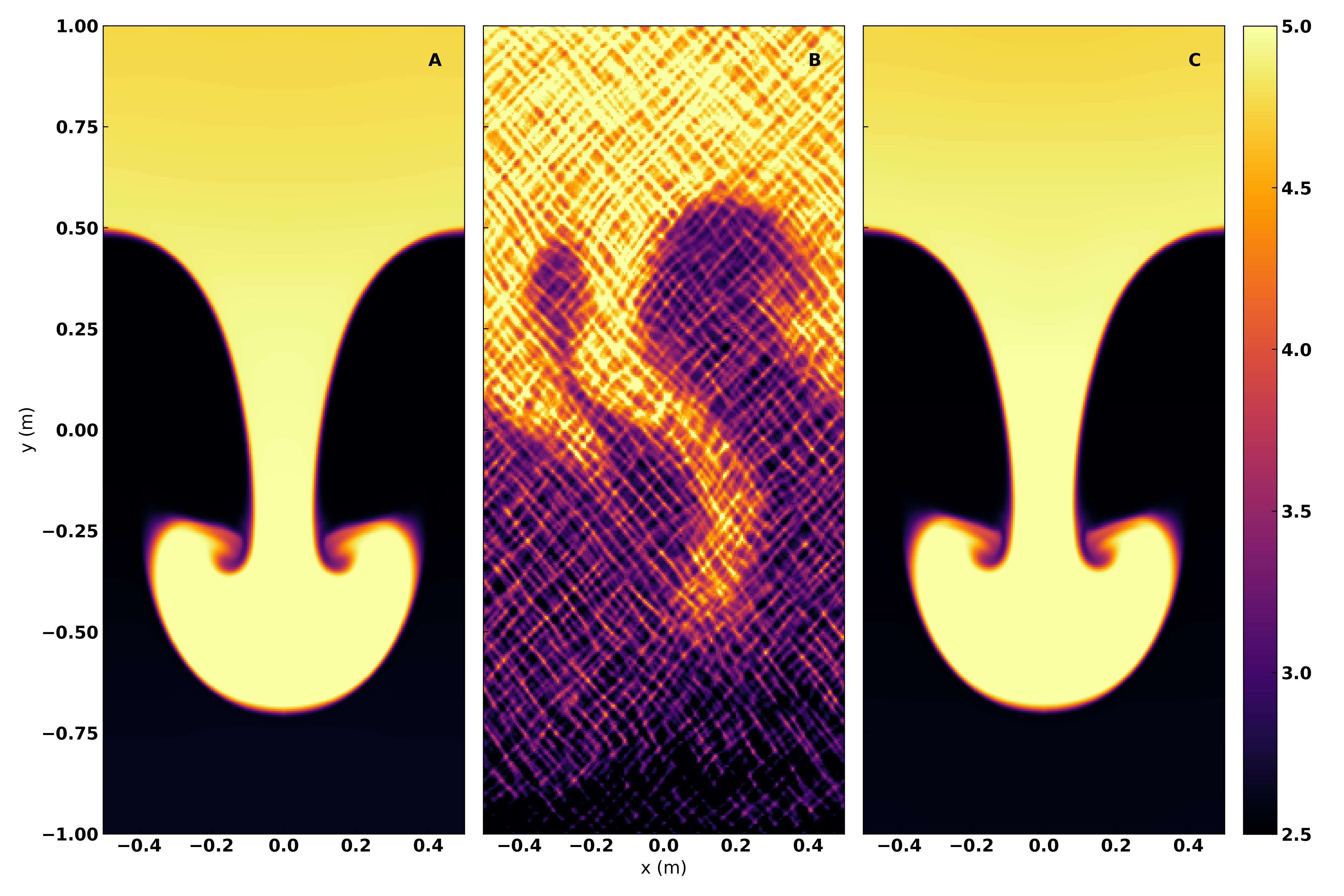}
  \caption{The final density state in kg/m$^3$ ($t= 27.5\;ms$) of the 2D MRT
  instability growth of density in kg/m$^3$ for the (A) ideal-gas EOS, (B) the
  SESAME-5760 EOS using the cell-centered EOS evaluation algorithm, and (C) the
  SESAME 5760-EOS using the interface EOS evaluation algorithm.}
  \label{fig:rtLate}
\end{figure}

Figures~\ref{fig:rtEarly} and \ref{fig:rtLate} show the growth of the MRT
instability early and late in time, respectively. Examination of these figures
highlights the differences between the different EOS-evaluation algorithms.
Consistent with the results of Secs.~\ref{sec:lw1D}, \ref{sec:shockTube} and
\ref{sec:lw2D} which demonstrated potential issues with the cell-centered-EOS-evaluation 
algorithm, poor results are observed with the cell-centered-EOS-evaluation 
algorithm (B) relative to computations with an ideal-gas EOS (A) and
the interface-EOS-evaluation algorithm (C).  However, unlike the prior results
which produced somewhat qualitatively consistent results with the cell-centered-EOS-evaluation
algorithm, the MRT test shows that computations with this
algorithm can be completely unreliable.  As seen in Fig.~\ref{fig:rtEarly}(B),
the cell-centered-EOS-evaluation algorithm generates spurious sound waves
almost immediately. By the end of the simulation, Fig.~\ref{fig:rtLate}(B),
this leads to unphysical results.  The results of the ideal gas model (A) are
qualitatively identical to the use of the interface-EOS-evaluation algorithm
(C) with only small differences in magnitude at the simulation end while the
propagation distance and mode number are the same.  As the linear-MRT analysis
is not dependent on the EOS, this indicates that the interface-EOS-evaluation
algorithm is convincingly more accurate than the cell-centered-EOS-evaluation
algorithm.

\section{Discussion and Summary}
\label{sec:summary}

This work evaluates algorithms that model applications described by
magnetohydrodynamics with a real-gas Equation-of-State in multiple dimensions.
The approximate Riemann solvers employed take the HLL form and two methods of
evaluation of the Equation-of-State are proposed: at the cell centers or
directly at the interfaces between cells. Although the former method is
computationally more efficient that the latter, we demonstrate that such an
approach lacks robustness through a suite of one- and two-dimensional tests.
For example, the cell-centered-EOS-evaluation algorithm exhibits poor
convergence with the linear magnetosonic waves in both 1D and 2D. In addition
to this failure, the cell-centered-EOS-evaluation algorithm produces spurious
results for 1D shock tube test problems at high resolution using an HLLD-type
Riemann solver. We note that the shock tube test problems considered here probe
EOS that include \emph{non-convex} behavior, indicated through the fundamental
thermodynamic derivative passing from positive to negative \cite{Serna2014}.
These tests therefore demonstrate the ability of the scheme to capture behavior
that is expected to occur in tabular EOS associated with real materials. By
contrast, the interface-EOS-evaluation algorithm produces high quality results
for this problem set at high resolution with the HLLD Riemann solver. Overall,
results for these shock tubes indicate that the interface-EOS-evaluation
algorithm may be more robust for tabular EOS associated with real materials,
even when combined with Riemann-solvers that minimize numerical dissipation.

That the cell-centered-EOS-evaluation algorithm lacks robustness for real EOS
is demonstrated by a 2D magneto-Rayleigh-Taylor instability test problem, where
this algorithm completely fails to produce reasonable results. These are
characterized by a checker board pattern that exhibits in the density at early
times during the linear growth phase of the instability and \emph{completely}
corrupts the development of the characteristic MRT fingers that characterize
the instability. By contrast, the interface-EOS-evaluation algorithm is able to
accurately capture the linear growth of this  2D magneto-Rayleigh-Taylor
instability, demonstrating the growth of the characteristic MRT fingers in a
fashion at least qualitatively comparable to the development of the instability
in an ideal gas. Since the linear growth phase of the instability is
independent of the EOS, the qualitative similarity of the results produced by
the scheme for both ideal and real gas (tabular) EOS serves to enhance
confidence in the interface-EOS-evaluation algorithm, while demonstrating the
unsuitability of the cell-centered-EOS-evaluation scheme.

This work is a first step on the route to developing robust, peer-reviewed
algorithms for combining magnetohydrodynamics with EOS for real materials. Our
principle conclusion is that the EOS must be evaluated collocated with points
for computation of numerical fluxes in order to produce a robust algorithm,
capable of accurately capturing MHD effects in real materials. As such, our
conclusions have important implications for the design of schemes that combine
higher order finite volume and/or discontinuous Galerkin methods for MHD with
real EOS. However, in demonstrating this conclusion, we have considered a
single material; however, real experiments often consist of multiple materials,
each with an equation of state that may exhibit different behavior. A range of
methodologies for mixing together EOS for different materials exist
\cite{Magyar2014}, which amount, in general to different rules for mixing
together material pressures in either either a linear or non-linear fashion
dependent on the fractional densities. We note that Ref.~\cite{Magyar2014}
found that a mixing model based on the (non-linear) Amagat's rule works
reliably in the multi Mbar and several thousand degree temperature range that
characterizes high energy density environments. In principle, such a rule can
be combined with algorithms that track species mass fractions (such as those
considered by, e.g. Ref.~\cite{Hu2009}) in order to develop an algorithm that
enables mixed materials with real EOS, which can, in turn, be combined with the
MHD algorithms developed here. Care, however, is needed, to ensure that such an
algorithm appropriately captures the convexity of the system when combined with
magnetic fields, as discussed by Ref.~\cite{Serna2014} and shown to be the 
case for the algorithms here in Sec.~\ref{sec:shockTube}. Furthermore, high
energy density systems often exist in strong coupling regimes, where the ionic
Coulomb interaction dominates over the ion thermal energy \cite{Stanton2015}.
In such systems, it is challenging to separate the EOS from the Coulomb
interaction and develop fluid models that sit on a rigorous theoretical
foundation \cite{Diaw2019,Baalrud2019}; the case including magnetic fields,
which, in the fluid limit, would yield MHD models that
consistently incorporate real materials EOS and strong coupling appropriate for
high energy density plasmas is, as yet, an unsolved problem. We leave detailed
consideration of all of these issues to future work.
\vspace{-0.200in}
\section*{Acknowledgements}
\vspace{-0.100in}

The authors would like to thank Peter Stoltz, Eric Held, Scott Kruger, Thomas
Mattsson, John Luginsland and an anonymous referee for helpful comments that
improved the quality of this manuscript.  This material is based on work
supported by the U.S.  Department of Energy Office of Science DE-SC0016531
(King) and DE-SC0016515 (Srinivasan and Masti). This paper describes objective
technical results and analysis. Any subjective views or opinions that might be
expressed in the paper do not necessarily represent the views of the U.S.
Department of Energy or the United States Government. Sandia National
Laboratories is a multimission laboratory managed and operated by National
Technology \& Engineering Solutions of Sandia, LLC, a wholly owned subsidiary
of Honeywell International Inc., for the U.S. Department of Energy's National
Nuclear Security Administration under contract DE-NA0003525 (support for
Beckwith). SAND Number: SAND2020-1197 J

\bibliographystyle{apsrev4-1}
\bibliography{Biblio}

\end{document}